\documentclass{article}
\usepackage[preprint]{neurips_2025}

\usepackage{url}  
\usepackage{cite}
\usepackage{amsmath,amssymb,amsfonts}
\usepackage{algorithmic}
\usepackage{graphicx}
\usepackage{textcomp}
\usepackage{xcolor}

\usepackage{balance}
\usepackage{listings}
\usepackage{comment}
\usepackage{tcolorbox}
\usepackage[ruled,vlined,linesnumbered]{algorithm2e}
\usepackage{multirow}
\usepackage{booktabs}
\usepackage{adjustbox}
\usepackage{threeparttable}

\definecolor{green}{RGB}{72,160,76}
\definecolor{red}{RGB}{172,53,44}

\newcommand\yc[1]{#1}

\newcommand\tian[1]{#1}

\newcommand{\tool}{GSE}

\newcommand{\nograph}{\tool$_{w/o\text{ }graph}$}
\newcommand{\nogen}{\tool$_{w/o\text{ }gen}$}

\renewcommand{\cite}{\citep}

\title{Learning Globally Reusable Skills for Coding Agents}

\author{
    Chen Yang$^{1}$\;
    Jiashuo Tian$^{1}$\;
    Ziqi Wang$^{1}$\;
    Xinyin Liu$^{1}$\;
    Meiru Ye$^{2}$\;
    Junjie Chen$^{1}$
    \\[\medskipamount]
    $^{1}$ Tianjin University \; 
    $^{2}$ Tsinghua University
    \\[\medskipamount]
    \texttt{\{yangchenyc, tianjiashuo, wangziqi123, liuxinyin, junjiechen\}@tju.edu.cn} \\
    \texttt{yemr22@mails.tsinghua.edu.cn}
}

\begin{document}

\maketitle

\frenchspacing

\begin{abstract}
Automated skill evolution enables Large Language Model (LLM) agents to continuously improve without expensive retraining. However, existing approaches typically treat skill evolution as a sequence of local updates, overlooking relationships among skills and often producing overfitted skill updates that fail to generalize across tasks. 
We propose \tool{}, a globalized skill evolution framework that jointly optimizes skill compatibility and skill generalization. 
To preserve consistency across the skill bank, \tool{} maintains a Skill Relation Graph (SRG) that explicitly models and co-evolves inter-skill relationships. To improve generalization, \tool{} performs cluster-based skill consolidation to abstract reusable capabilities from local updates and employs replay-driven verification to prevent overfitting and behavioral regressions. 
We evaluate \tool{} on two representative software engineering tasks: bug-revealing test generation and false-positive bug report filtering. Across two state-of-the-art coding agents, OpenHands and mini-SWE-agent, \tool{} consistently achieves the best precision, recall, and F1-score. 
\tian{Compared with existing evolution techniques,  }\tool{} improves precision and recall by \tian{6.1\%$\sim$34.1\%} and \tian{31.8\%$\sim$180.0\%} for test generation, and by \tian{15.4\%$\sim$96.4\%} and \tian{13.1\%$\sim$19.8\%} for false-positive filtering.
Deployment on an internal industrial agent further yields a \tian{61.4\%} improvement in F1-score, demonstrating the effectiveness and generalizability of \tool{} for evolving effective skills.

\end{abstract}

\section{Introduction}
\label{sec:intro}

Large Language Model (LLM) agents are increasingly used to automate a wide range of software engineering tasks, such as test generation and defect analysis.
By navigating complex repositories, executing commands, and iteratively refining their actions based on environment feedback, these agents offer a promising path toward autonomous defect discovery and validation in large-scale software systems.

Despite their strong performance, such agents are often limited by static designs. Fixed prompts, tool policies, and manually curated skills can quickly become insufficient as coding tasks vary across programming languages, project conventions, and failure modes. 
To improve agents without the high cost of model retraining, automatic agent evolution has attracted increasing attention~\cite{agent_evolve_survey,live_swe_agent,trace2skill}. 
Among evolution techniques, \textit{skill evolution} is particularly appealing: it extracts and refines reusable abstractions of reasoning and action patterns from execution histories, enabling agents to continuously adapt their behavior through very lightweight updates.

Recent automated skill evolution techniques show encouraging results~\cite{trace2skill,codeskill}.
Typically, these techniques first analyze individual execution traces to generate skill updates (i.e., create new skills or revise existing ones), and then integrate the resulting skills into an agent’s reusable skill bank.
However, they largely treat skill evolution as a sequence of local updates, whereas both the generation and integration stages fundamentally require a global perspective. At the update generation stage, an update derived solely from isolated traces may violate constraints assumed by other existing skills. At the integration stage, a locally successful update may be overly specialized to the observed traces and can introduce regressions when reused in future tasks.
As the skill bank grows, such locally optimized updates can accumulate and gradually degrade the overall quality and coherence of the agent’s skill set. These effects are not immediately apparent from individual updates, but emerge from their interactions within the evolving system.
Therefore, \textit{skill evolution should be formulated as a global optimization problem rather than a sequence of independent local updates}. However, achieving globally effective skill evolution introduces two key challenges.

\textbf{Challenge 1: How to generate skill updates that are compatible with an existing skill bank?}
A coding agent operates as a tightly coupled system in which the output of one skill often serves as the input assumption of another. For example, a context-exploration skill determines which code contexts are considered relevant, while a bug-reasoning skill constructs hypotheses based on the selected context.
When a skill is updated in isolation to address a specific failure, it may inadvertently alter assumptions relied upon by other skills. Moreover, skills may exhibit complementary or conflicting relationships, yet such relationships are rarely represented explicitly. Without a mechanism to model and manage these inter-skill relationships, skill evolution may gradually lead to incompatible skill behaviors within the bank.

\textbf{Challenge 2: How to merge locally successful updates into a globally effective skill bank?}
Skill updates derived from individual execution traces are often highly specific to the contexts in which they were generated. Furthermore, different traces may independently yield updates that reflect the same underlying capability gap or behavioral pattern, yet encode them through different low-level observations or representations.
Directly merging such updates risks introducing overly specific and fragmented skills into the skill bank, which can prevent the agent from learning more general and reusable capabilities. As a result, the agent may overfit to particular projects, bug patterns, or execution scenarios. Without mechanisms to abstract reusable skill representations and to verify that updates generalize beyond the original traces, the skill bank may gradually accumulate fragmented and overfitted skills, ultimately degrading long-term effectiveness.

To address these challenges, we propose \textbf{\tool{}}, a \textbf{G}lobalized \textbf{S}kill \textbf{E}volution framework that optimizes the skill bank as an interconnected and reusable system, rather than applying isolated updates.
To address the first challenge, \tool{} generates skill evolution proposals that explicitly account for relationships among skills via a Skill Relation Graph (SRG). Given an execution trace, an update agent first identifies the capability deficiencies responsible for the observed failure and then formulates a structured skill evolution proposal in a domain-specific language (DSL), specifying the \textit{operation}, \textit{target skill}, \textit{content}, \textit{rationale}, and \textit{expected effect}.
To enable globally consistent evolution, \tool{} maintains an SRG that explicitly models inter-skill relationships, including \textit{dependency}, \textit{co-usage}, and \textit{conflict}. This graph provides a global view of the skill bank, enabling reasoning about how a proposed update may propagate across related skills.
Guided by this global dependency structure, \tool{} jointly evolves both skill contents and their relationships. When a proposed update affects related skills, \tool{} generates corresponding auxiliary evolution proposals to preserve consistency and compatibility across the skill bank.

To address the second challenge, \tool{} introduces a global skill generalization mechanism that transforms trace-specific updates into reusable skill abstractions via cluster-based consolidation and replay-driven verification.
Given a proposal, \tool{} first applies the update and replays the originating trace to ensure it resolves the triggering failure, providing local correctness validation. Validated proposals are then clustered based on affected skills and shared rationales, enabling the identification of common capability patterns across traces.
Within each cluster, \tool{} aggregates case-specific updates into unified, higher-level skills, reducing redundancy and mitigating overfitting to specific execution contexts. The resulting skills are further evaluated through replay on historical cases to assess robustness and detect regressions. Only skills that consistently pass this global verification are integrated into the skill bank.
Through this process, \tool{} converts fragmented trace-level improvements into coherent and reusable skill representations, enabling globally effective skill evolution.

We evaluate \tool{} on two representative software engineering tasks: \textit{bug-revealing test generation} and \textit{false-positive bug report filtering}. 
For test generation, we construct a Java benchmark derived from \textbf{Multi-SWE-Bench}~\cite{multiswebench}, comprising 108 real-world bugs from 9 open-source projects. For false-positive filtering, we collaborate with a leading international IT company (i.e., ByteDance)
to build \textbf{IndustrialBugs}, an industrial dataset containing 500 bug reports from 8 production repositories, including 132 confirmed bugs and 368 false alarms.
We instantiate two state-of-the-art coding agents, OpenHands~\cite{openhands} and mini-SWE-agent~\cite{swe_agent}, as base systems. For each, we compare five settings: the vanilla agent, the agent augmented with human-written skills, Live-SWE-agent~\cite{live_swe_agent}, Trace2Skill~\cite{trace2skill}, and \tool{}.
\yc{Both Live-SWE-agent and Trace2Skill represent the state-of-the-art evolution techniques.}
Results show that \tool{} consistently outperforms all baselines across precision, recall, and F1 on both tasks. In particular, it improves precision by \tian{6.1\%$\sim$34.1\%} and recall by \tian{31.8\%$\sim$180.0\%} on test generation, and achieves gains of \tian{15.4\%$\sim$96.4\%} in precision and \tian{13.1\%$\sim$19.8\%} in recall on false-positive filtering. Ablation studies further confirm that both the skill relation graph and the global skill generalization are critical to overall performance.

In summary, our contributions are as follows:

\begin{itemize}
\item We propose \tool{}, a globalized skill evolution framework that formulates skill evolution as a global optimization problem over an interconnected and reusable skill bank.

\item \tool{} introduces a Skill Relation Graph to explicitly model and co-evolve inter-skill relationships, together with a skill generalization mechanism that abstracts reusable capabilities while mitigating overfitting and preventing behavioral regressions.

\item We evaluate \tool{} on two representative software engineering tasks using both open-source and industrial datasets. Experimental results show that \tool{} consistently outperforms base agents, human-expert-written skills, and existing skill evolution techniques.

\end{itemize}

\section{Background}
\label{sec:background}
To contextualize the design of \tool{}, we introduce skill-augmented agent execution and automated skill evolution.

\subsection{Skill-Augmented Agent Execution}
Consider an autonomous agent assigned a software engineering task $x \in \mathcal{X}$ (e.g., generating a bug-revealing test or analyzing a bug report). The agent tackles the task by interacting with a code repository, producing a sequence of reasoning steps and tool invocations denoted as an execution trace $\mathcal{T}$. A validation function $V(\mathcal{T}, x) \in \{0, 1\}$ evaluates whether the trace successfully resolves the task.

While modern LLM agents can perform direct reasoning, their zero-shot capabilities are often insufficient for domain-specific or complex repositories. To reduce redundant exploration and enforce systematic workflows, agents are augmented with a \textit{skill bank} $\mathcal{D} = \{s_1, \dots, s_{|\mathcal{D}|}\}$. 
A skill $s \in \mathcal{D}$ typically encapsulates reusable procedural knowledge, tool schemas, and environment constraints. 
For a given task $x$, a retrieval mechanism fetches a relevant subset of skills $\mathcal{S}_x \subseteq \mathcal{D}$. 
The agent then generates its trajectory by conditioning its policy on both the task and the retrieved skills:
\begin{equation}
\tau \sim \pi(\cdot \mid \mathcal{S}_x, x), x \in \mathcal{X}.
\end{equation}
Consequently, the agent's expected success rate, defined as $\mathbb{E}_{x \sim \mathcal{X}}[V(\pi(\cdot \mid x,\mathcal{D}), x)]$, depends directly on the quality of the underlying skill bank $\mathcal{D}$.

\begin{figure*}[t]
  \centering
  \includegraphics[width=\linewidth]{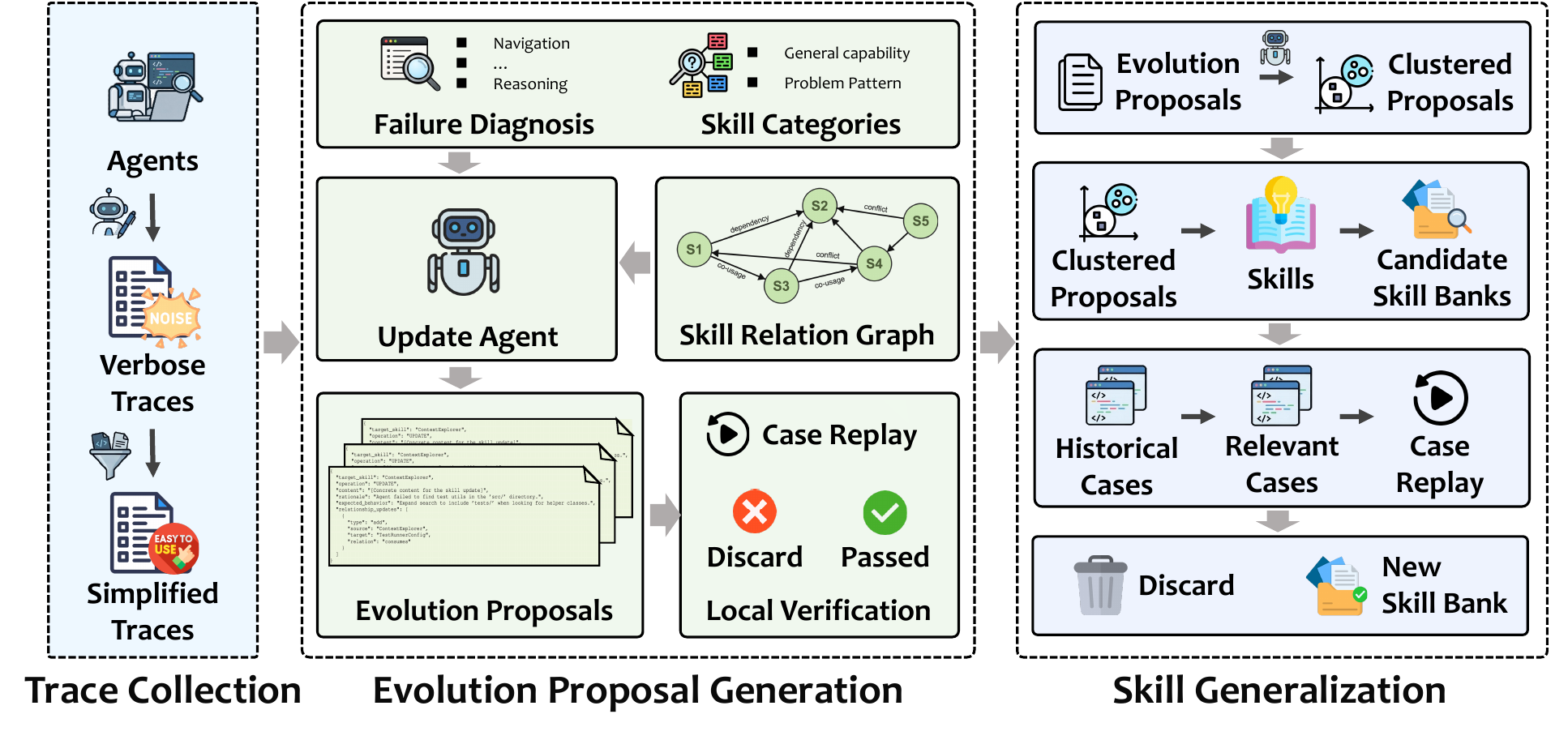}
  \caption{Overview of \tool{}}
  \label{fig:overview}
  \vspace{-10pt}
\end{figure*}

\subsection{Automated Skill Evolution}
As software systems and task requirements change, static skill banks inevitably become outdated or insufficient. 
\textit{Automated skill evolution} addresses this limitation by refining the skill bank based on the agent's own execution history. 

The evolution paradigm operates as a continuous feedback loop between execution and reflection. 
Given an execution trajectory $\tau$, where $V(\tau, x)$ evaluates the outcome of solving task instance $x$, an evolution module $\mathcal{U}$ analyzes the trajectory and proposes updates to the skill bank, producing an evolved skill bank $\mathcal{D}'$:
\begin{equation}
\mathcal{D}' = \mathcal{U}(\mathcal{D}, \tau).
\end{equation}
The objective of skill evolution is to find an optimal skill bank $\mathcal{D}^*$ that maximizes the expected success rate across the distribution of all tasks.

\section{Approach}
\label{sec:approach}

Figure~\ref{fig:overview} presents the workflow of \tool{}, which optimizes the skill bank through relationship-aware evolution proposals and skill generalization. 
First, \textbf{trace collection} captures raw execution traces and distills them into semantic-rich traces that preserve the agent's reasoning, actions, and task outcomes. 
Next, in the \textbf{evolution proposal generation} stage, an update agent identifies the underlying capability deficiency and formulates a structured skill evolution proposal. To preserve skill compatibility, \tool{} maintains a Skill Relation Graph (SRG) that explicitly models and co-evolves inter-skill relationships. 
Finally, in the \textbf{skill generalization} stage, \tool{} aggregates locally effective proposals into more general and reusable skills and validates them through replay on relevant historical cases to prevent overfitting and behavioral regressions before merging them into the skill bank.

\subsection{Trace Collection}
\label{sec:basic_info}

Before learning reusable skills from agent failures, we must accurately capture and comprehend its execution history.
Therefore, \tool{} first instruments the underlying agent to record all actions, including the agent's internal reasoning steps (e.g., ``reasoning'' blocks), invoked tools, command parameters, environment feedback, and the final task outcome.
However, raw execution traces are often extremely long, containing verbose outputs from file-system inspections, repetitive compilation errors, and intermediate tool responses. Directly feeding raw traces into an evolution module is not only computationally expensive, but it also obscures the analysis process.
Therefore, \tool{} employs trace simplification to distill the semantic-rich trace needed for subsequent analysis.

Specifically, trace simplification filters out token-heavy command outputs that do not contribute to high-level causal analysis.
It removes verbose outputs from file-search commands, such as the full results of \texttt{find} or \texttt{grep}, raw file contents returned by file-operating tools, and low-value informational logs (i.e., info-level logs) produced by compilation or test commands.
At the same time, it preserves the semantic metadata of these actions, including the exact commands executed, the targeted file paths, key error or warning messages, and the agent's internal reasoning blocks.
By discarding noisy content while preserving the sequential logic and causality of the agent actions, the simplified trace enables the subsequent evolution module to efficiently perform analysis without being overwhelmed by low-level execution details.

\subsection{Evolution Proposal Generation}
\label{sec:update_agent}

Given the simplified trace of an agent failure, the next step is to determine how the skill bank should evolve to prevent similar failures in the future. 
Existing skill evolution methods typically derive updates from individual traces and apply them independently, overlooking relationships among skills and potentially introducing incompatible skills into the skill bank. 
To address this limitation, \tool{} employs an \textit{update agent} that diagnoses the underlying capability deficiency, reasons about inter-skill relationships through the Skill Relation Graph (SRG), and generates structured evolution proposals that preserve the compatibility of the overall skill bank.

\subsubsection{Failure Diagnosis}
To support effective skill evolution, the update agent first identifies the capability deficiency underlying the observed failure.
Specifically, it performs a hierarchical diagnosis to determine whether the failure originates from deficient context exploration (navigation failure), incorrect reasoning about the codebase (reasoning failure), or improper execution behavior (execution failure).
For example, the agent may fail because it overlooked relevant files due to an overly narrow search strategy, or because it identified the bug correctly but generated a generic test that failed to trigger it.
Based on this diagnosis, the update agent formulates a root-cause hypothesis describing the missing capability.

The diagnosed capability deficiency then determines the direction of skill evolution.
\tool{} distinguishes two complementary forms of skill evolution.
General capability enhancement improves high-level behaviors such as context exploration and validation strategies.
This type of evolution is triggered when the diagnosed deficiency reveals limitations in the agent's workflow or decision-making process.
Reusable problem pattern captures recurring code or task characteristics, such as bug-triggering patterns, library constraints, or framework-specific strategies.
This type of evolution is triggered when the diagnosed deficiency reflects an unrecognized but reusable pattern.
Each pattern skill specifies both the characteristic pattern and an actionable strategy for handling it.

\subsubsection{Skill Relation Graph (SRG)}
A key challenge in skill evolution is that skills do not operate independently.
The output assumptions, execution order, and applicability conditions of one skill often influence the behavior of others. 
Therefore, \tool{} models the skill bank as a \textbf{Skill Relation Graph (SRG)}, which explicitly represents and evolves the dependencies among skills.
Formally, the SRG is a directed graph $G=(V,E)$, where each node $v\in V$ represents a skill in the agent's skill bank, and each edge $e=(v_i,v_j,r)\in E$ represents a semantic relationship of type $r\in\mathcal{R}$ between two skills. Unlike prior work that treats skills as independent artifacts, \tool{} treats both the skill contents and their relationships as first-class objects of evolution.
The relationship set $\mathcal{R}$ includes three types:
\begin{itemize}
\item \texttt{dependency}: skill $v_j$ relies on assumptions, or results produced by skill $v_i$. This relationship captures execution and reasoning dependencies among skills.

\item \texttt{co-usage}: skills $v_i$ and $v_j$ are complementary and typically achieve better performance when applied together or in a particular order.

\item \texttt{conflict}: skills $v_i$ and $v_j$ provide incompatible guidance or encode contradictory assumptions, and therefore should not be activated simultaneously.

\end{itemize}

The SRG enables \tool{} to reason about the global impact of a local evolution proposal.
Given a proposed skill update, \tool{} first identifies potentially affected skills by examining all neighboring skills connected to the updated skill in the SRG.
To account for interactions that are not yet explicitly represented, \tool{} further retrieves candidate related skills from the entire skill bank 
using skill metadata, such as capability category, task type, and execution stage, and then performs semantic analysis of their contents to identify potential interactions.
The update agent then determines whether the proposed modification changes the assumptions, intermediate outputs, applicability conditions, or interaction patterns of each affected skill.
If so, \tool{} generates corresponding auxiliary evolution proposals that jointly evolve both the affected skills and their relationships.
By jointly evolving skill contents and inter-skill relationships through relationship-aware propagation, \tool{} transforms skill evolution from a sequence of independent local updates into a globalized optimization process over the entire skill bank.

\begin{figure}[t]
    \centering
    \includegraphics[width=0.85\linewidth]{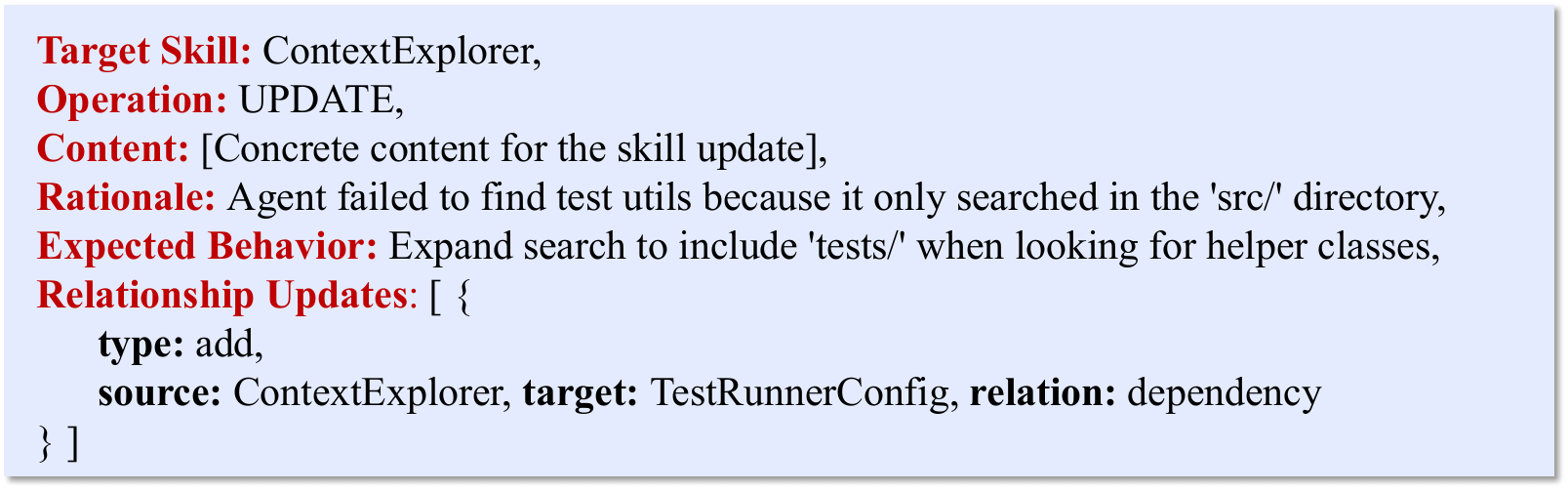}
    \label{fig:dsl_example}
    \caption{An Example of the Evolution Proposal DSL}
\end{figure}

\subsubsection{DSL-Based Evolution Proposal}
To reduces ambiguity and facilitate downstream processing, the evolution proposal is strictly structured using a specialized Domain-Specific Language (DSL).
Concretely, it defines a standardized schema for modifying both skill nodes and skill relations.
The DSL specifies the target skill, the edit operation, the concrete content to be updated, the rationale, the expected behavior after the update, and the relation changes. Figure~\ref{fig:dsl_example} shows one example.

Before performing skill generalization, \tool{} first verifies the local effectiveness of each evolution proposal.
Specifically, \tool{} applies the proposed changes to a temporary candidate skill bank and replays the failure case that produced the proposal.
Proposals that fail to resolve their originating failure are discarded immediately.
This local validation step filters out ineffective proposals early and avoids unnecessary generalization costs.

\subsection{Skill Generalization}
\label{sec:quality_gate}

Even after local validation, transforming case-specific evolution proposals into reusable skills that improve the skill bank globally remains challenging.
Proposals derived from individual failures often encode observations specific to particular projects, bug patterns, or execution scenarios, and directly merging them may lead to fragmented and overfitted skills.
To address this challenge, \tool{} performs \textit{skill generalization} through two stages: \textit{cluster-based skill consolidation} and \textit{replay-driven validation}.
Algorithm~\ref{alg:skill_generalization} summarizes the process.

\begin{algorithm}[t]
\caption{Skill Generalization Process}
\label{alg:skill_generalization}
\begin{algorithmic}[1]
\REQUIRE Locally validated evolution proposals $\mathcal{P}$, current SRG $G=(V,E)$, historical case set $\mathcal{H}$

\STATE $\mathcal{C}_{exist} \leftarrow \emptyset$, $\mathcal{C}_{new} \leftarrow \emptyset$, $\mathcal{N} \leftarrow \emptyset$

\COMMENT{Cluster evolution proposals}
\FOR{each proposal $p \in \mathcal{P}$}
    \STATE $S_p \leftarrow$ existing skills modified by $p$
    \STATE $N_p \leftarrow$ new skills introduced by $p$

    \IF{$S_p \neq \emptyset$}
        \STATE Group $p$ into $\mathcal{C}_{exist}$ using $S_p$ as the clustering key
    \ENDIF

    \IF{$N_p \neq \emptyset$}
        \STATE Add $(p,N_p)$ to the new-skill pool $\mathcal{N}$
    \ENDIF
\ENDFOR

\STATE $\mathcal{C}_{new} \leftarrow$ cluster $\mathcal{N}$ by semantic similarity

\COMMENT{Generalize and validate each cluster}
\FOR{each cluster $c \in \mathcal{C}_{exist} \cup \mathcal{C}_{new}$}

    \STATE $a_c \leftarrow$ extract common capability patterns from $c$
    \STATE $G_c \leftarrow$ instantiate $a_c$ on $G$
    
    \IF{$c \in \mathcal{C}_{exist}$}
        \STATE $R_c \leftarrow$ historical cases using the affected skills
    \ELSE
        \STATE $R_c \leftarrow$ cases that generated proposals in $c$
    \ENDIF

    \STATE $o_c \leftarrow$ replay $R_c$ using candidate skill bank $G_c$

    \IF{$o_c$ preserves or improves performance}
        \STATE $G \leftarrow G_c$
    \ENDIF

\ENDFOR

\RETURN $G$
\end{algorithmic}
\end{algorithm}

\subsubsection{Skill Consolidation}
The goal of skill consolidation is to identify evolution proposals that reflect the same underlying capability deficiency and abstract their shared behaviors into more general and reusable skills. 
To achieve this, \tool{} groups locally effective evolution proposals into clusters and extracts the common capability patterns represented by each cluster.
For proposals that modify \textit{existing skills}, \tool{} uses the affected skill set as the clustering key. The intuition is that failures requiring modifications to the same skills are likely to expose similar capability deficiencies and therefore should be generalized jointly. 
For proposals that introduce \textit{new skills}, no corresponding nodes exist in the SRG. Therefore, \tool{} clusters them according to semantic similarity. 
Specifically, \tool{} compares the rationale and expected behavioral effects of the proposed skills to identify shared capability gaps.

After clustering, a merge agent analyzes the proposals within each cluster to extract their common patterns, remove case-specific details, and resolve redundant or conflicting updates. 
The resulting skill abstractions are then applied to construct a \emph{candidate skill bank}, which represents a generalized version of the clustered evolution proposals and serves as the input to the subsequent replay-driven validation stage.

\subsubsection{Replay-Driven Validation}
Although skill consolidation reduces fragmentation and overfitting, the resulting generalized skills may still fail to transfer beyond the cases from which they were derived.
Therefore, \tool{} performs replay-driven validation to evaluate whether the generalized skills truly capture reusable capabilities.
For each candidate skill bank, \tool{} replays a set of relevant historical cases.
For those that modify existing skills, \tool{} retrieves historical cases in which the affected skills were previously activated, since these cases are most likely to expose regressions or compatibility issues.
For those that introduce new skills, \tool{} uses all cases that generated the clustered proposals as the replay set.
A candidate skill bank is accepted only if it maintains or improves performance across the replay cases.
This validation process ensures that the resulting skills generalize beyond individual failures and prevents overfitted or regressive updates from entering the skill bank.

\section{Evaluation Design}
\label{sec:setup}

\yc{We evaluate \tool{} on two representative software engineering tasks: (i) \textit{bug-revealing test generation} and (ii) \textit{false-positive bug report filtering}. 
These tasks exercise many of the core capabilities required by software engineering agents in general, including repository exploration, code comprehension, semantic reasoning, constraint-aware code generation, program execution, and defect diagnosis. 
These capabilities are fundamental building blocks underlying a wide range of software engineering tasks, such as bug fixing, code review, and program analysis. 
Therefore, evaluating \tool{} on these two scenarios enables us to assess whether our skill evolution framework can learn effective skills across diverse capability requirements rather than overfitting to a single task.
}
Specifically, we formulate the following research questions (RQs):
\begin{itemize}
    \item \textbf{RQ1: To what extent can \tool{} support bug-revealing test generation?}
    The first goal of \tool{} is to evolve skills that help an agent generate unit tests capable of exposing real bugs in the focal method. 

    \item \textbf{RQ2: To what extent can \tool{} support false-positive bug report filtering?}
    The second goal of \tool{} is to evolve skills that help an agent decide whether bug reports produced by upstream tools correspond to real defects or false alarms. 

    \item \textbf{RQ3: How does each main component of \tool{} contribute to its overall effectiveness?}
    This RQ motivates an ablation study to evaluate the individual impact of \tool{}'s core components on its overall effectiveness.
    
    \item \textbf{RQ4: What is the cost of \tool{}, and how does it affect downstream execution efficiency?}
    This RQ evaluates the token cost of \tool{} and its impact on downstream task execution efficiency.
\end{itemize}

\subsection{Datasets}
\label{sec:dataset}
\subsubsection{Bug-Revealing Test Generation}
For the test generation task, we build our benchmark on top of \textbf{Multi-SWE-Bench}~\cite{multiswebench}, a large-scale repository-level dataset of real-world bugs and developer-authored fixes.
We focus on its Java portion because Java is one of the most widely-used languages in both industry and open-source communities.
For each bug, the dataset provides a developer fix patch.
We use abstract syntax tree (AST) parsing to identify the methods modified by each patch and treat them as candidate focal methods.
For each candidate, we manually verify whether invoking the method can trigger the corresponding bug, and retain only verified triggerable methods as buggy methods.
When a bug involves multiple such methods, we randomly select one as the focal method.
This procedure yields 108 buggy focal methods spanning 9 actively maintained Java projects.
The focal methods are non-trivial with 4$\sim$206 lines of code, on average 6 branches, 8 direct callees, and references to symbols defined in 5 external files (e.g., classes, interfaces, helper utilities), reflecting the complexity.

\subsubsection{False-Positive Bug Report Filtering}
For this task, we collaborate with our industrial partner, an internationally leading IT and AI technology company, to construct \textbf{IndustrialBugs}, a \textbf{diverse} dataset of bug reports produced by their internally-developed code-review tool for Go.
The diversity is reflected by the following aspects:
(1) The reports are collected across 8 production code repositories spanning 3 internal business lines and amount to 500 reports in total, of which 132 are confirmed real bugs and 368 are false alarms, with every report inspected and labeled by the original committer.
(2) The repositories range from 12.8$\sim$186.3 KLoC and contain 23$\sim$444 Go source files, and cover diverse domains including inter-service communication components (e.g., RPC-based microservices), external-facing interfaces (e.g., API implementations), and core production services supporting critical business workflows, reflecting typical industrial diversity in implementation style and runtime stack.
(3) The reports also cover diverse bug types, including security-related issues (e.g., inconsistencies in authentication logic and improper handling of sensitive credentials), concurrency errors (e.g., deadlocks from multi-service interactions), data-flow and integration bugs (e.g., incorrect propagation or trans- formation of inputs across services), and functional correctness bugs (e.g., partially or incorrectly implemented functionalities).
(4) The false alarms are also diverse, arising from missed framework or service-level contracts, incomplete interprocedural reasoning, and benign patterns introduced by engineering conventions, defensive checks, or environment-specific assumptions.

\yc{We consider the two tasks representative for evaluating skill evolution since they jointly cover the core capabilities required by software engineering agents.
}
For both tasks, we adopt a one-project-held-out evaluation protocol.
Each case corresponds to a focal method for test generation or a bug report for false-positive filtering.
In each run, all cases from one project are held out for evaluation, 
\yc{ while cases from the remaining projects are used to evolve the skill bank.}
This protocol ensures that no evaluated focal method or bug report is used during evolution and tests whether the evolved skills generalize across projects.

\subsection{Base Agents}
To reduce confounding effects from agent harnesses, 
\yc{we instantiate all compared techniques on two state-of-the-art coding agents (i.e., OpenHands~\cite{openhands} and mini-SWE-agent~\cite{swe_agent}) according to the SWE-bench leaderboard~\cite{swe_leaderboard}.}
These agents serve as the underlying base harnesses on which different \yc{techniques} are applied and evaluated.
Since mini-SWE-agent does not natively support skills, we extend it with a lightweight skill-loading interface so that all skill-based methods can be evaluated under the same setting.
For each base agent, all compared techniques use the same tools, execution environment, base LLM, and per-task budget.
This design ensures that performance differences mainly reflect the \yc{compared techniques} rather than differences in the underlying agent harnesses.

\subsection{Compared Techniques}
\yc{
We compare \tool{} with representative approaches for creating and evolving skills in coding agents.
}

\begin{itemize}

\item 
\yc{
\textbf{Human Skills}: A manually constructed skill set. We surveyed multiple public skill sources, including Anthropic Skills~\cite{anthropic_skills}, SkillsBench~\cite{skillsbench}, and SWE-Skills-Bench~\cite{sweskills}, but found that their coding-related skills mainly target general library usage or software development tasks and are not directly applicable to test generation or bug report analysis. 
Therefore, we invited \textit{6} senior software-quality engineers from our partner company, each with \textit{6}$\sim$\textit{10} years of engineering experience, to independently author skills and then jointly discuss, polish, and consolidate them. This process produced \textit{17} skills covering both high-level workflows and bug-specific knowledge.
}

\item 
\yc{
\textbf{Live-SWE-agent}~\cite{liveswe}: A state-of-the-art evolution framework that automatically evolves software engineering agent scaffolds. 
Although originally designed to evolve tools, its lightweight and general evolution workflow can be naturally adapted to skill evolution. 
We therefore adapt it to generate and update skills. 
}

\item 
\yc{
\textbf{Trace2Skill}~\cite{trace2skill}: A state-of-the-art evolution framework that 
leverages multiple analyst sub-agents to propose skill updates and merges them at once into one skill bank,
}
\end{itemize}

\yc{
We instantiate each compared technique, as well as \tool{}, on two base agents (i.e., OpenHands~\cite{openhands} and mini-SWE-agent~\cite{swe_agent}). In addition, we include the corresponding \textbf{vanilla base agents}, which execute the tasks without any skills or evolution mechanisms, to characterize the capabilities of the underlying agents.
}

\subsection{Measurements}
\label{sec:metric}
We use precision, recall, and F1-score as the primary evaluation metrics for both tasks, following prior work on test generation~\cite{toga,rted} and bug triage~\cite{linux_fp, tencent_fp}.

\subsubsection{Outcomes for Bug-Revealing Test Generation}
For each focal method in the test set, we run the technique under evaluation and execute its generated tests. Following prior work~\cite{toga,rted}, we categorize the outcomes as:
\begin{itemize}
\item True Positive (TP$_{\textit{bug}}$): the test fails because of the ground-truth bug.
\item False Positive (FP$_{\textit{bug}}$): the test fails, but the failure is unrelated to the ground-truth bug.
\item False Negative (FN$_{\textit{bug}}$): the test does not reveal any bugs.
\end{itemize}
Since all focal methods in this task are known to be buggy, there are no true negatives.

\subsubsection{Outcomes for False-Positive Filtering}
For each report in the test set, we run the technique to predict whether the report describes a real bug or a false positive, and compare its prediction with the developer-confirmed label:
\begin{itemize}
\item True Positive (TP$_{\textit{filter}}$): the technique correctly identifies a real bug as a real bug.
\item True Negative (TN$_{\textit{filter}}$): the technique correctly identifies a false alarm as a false alarm.
\item False Positive (FP$_{\textit{filter}}$): the technique labels a false-alarm as a real bug.
\item False Negative (FN$_{\textit{filter}}$): the technique labels a real bug as a false alarm.
\end{itemize}

\subsubsection{Metric Calculation}
Based on these outcomes, we measured the effectiveness of each technique using the metrics:

\textbf{Precision} measures the proportion of true bugs among all samples identified as bugs: $\frac{\text{TP}}{\text{TP}+\text{FP}}$.
\textbf{Recall} measures the proportion of true bugs correctly identified:
$\frac{\text{TP}}{\text{ALL\_BUGS}}$.
For bug-revealing test generation, $\text{ALL\_BUGS}=\text{TP}_{\textit{bug}}+\text{FN}_{\textit{bug}}+\text{FP}_{\textit{bug}}$, where $\text{FP}_{\textit{bug}}$ denotes cases in which the generated test reveals a bug unrelated to the ground-truth bug.
For false-positive filtering, $\text{ALL\_BUGS}=\text{TP}_{\textit{filter}}+\text{FN}_{\textit{filter}}$, following the standard classification setting.
\textbf{F1-score} is the harmonic mean of Precision and Recall: $\frac{2 \times \text{Precision} \times \text{Recall}}{\text{Precision}+\text{Recall}}$.

\subsection{Implementation and Environment}
We implemented \tool{} in Python using LangChain (v1.1.1)~\cite{langchain}. All agents in \tool{} are powered by DeepSeek-V4-Flash via API services~\cite{deepseek}. Execution traces, replay outcomes, and update states are persisted in SQLite, while the skill bank is versioned with Git worktree. 
This design prevents concurrent evolution runs from interfering with one another and makes every accepted skill version reproducible.
For the base agents, we instantiate OpenHands~\cite{openhands} using its official implementation and extend the official mini-SWE-agent~\cite{swe_agent} with a lightweight skill-loading component.
For compared techniques (i.e., Live-SWE-agent and Trace2Skill), we reuse their released artifacts and adapt only their input/output interfaces so that they can operate on the same tasks as \tool{}.
All base agents and compared techniques use DeepSeek-V4-Flash as the underlying LLM via the same API service. 
All experiments were conducted on a workstation running Ubuntu 20.04, equipped with a 128-core CPU and 128 GB of RAM.

\section{Results and Analysis}
\label{sec:results}
\subsection{RQ1: Effectiveness on Bug-Revealing Test Generation}
\subsubsection{Process}
For each focal method, the agent generates a test, which is then executed to determine whether it exposes a bug.
We classify the outcomes using the definitions in Section~\ref{sec:metric} and report precision, recall, and F1-score.

\subsubsection{Results}
\begin{table}[t]
\centering
\caption{Effectiveness Comparison on Bug-Revealing Test Generation}
\label{tab:rq1}
\resizebox{0.75\columnwidth}{!}{
\begin{tabular}{l |ccc| ccc}
\toprule
\multirow{2}{*}{\textbf{Method}} & \multicolumn{3}{c|}{\textbf{OpenHands}} & \multicolumn{3}{c}{\textbf{mini-SWE-agent}} \\
& \textbf{Precision} & \textbf{Recall} & \textbf{F1} & \textbf{Precision} & \textbf{Recall} & \textbf{F1} \\
\midrule
Base Agent & 0.17 & 0.06 & 0.08 & 0.48 & 0.19 & 0.28 \\
+ Human Skill & 0.33 & 0.10 & 0.16 & 0.42 & 0.19 & 0.26 \\
+ Live-SWE-agent & 0.31 & 0.17 & 0.22 & 0.43 & 0.21 & 0.29 \\
+ Trace2Skill & 0.31 & 0.14 & 0.19 & 0.41 & 0.22 & 0.29 \\
\midrule
+ \tool{} & \textbf{0.35} & \textbf{0.28} & \textbf{0.31} & \textbf{0.55} & \textbf{0.29} & \textbf{0.38} \\
\bottomrule
\end{tabular}
}
\end{table}

Table~\ref{tab:rq1} presents the results.
From the table, \tool{} consistently outperforms all baselines on both OpenHands and mini-SWE-agent.
Specifically, on OpenHands, \tool{} obtains a precision of 0.35, recall of 0.28, and F1-score of 0.31, outperforming the strongest baseline scores of 0.33, 0.17, and 0.22 by 6.1\%, 64.7\%, and 40.9\%, respectively.
On mini-SWE-agent, \tool{} achieves a precision of 0.55, recall of 0.29, and F1-score of 0.38, improving over the strongest baselines by \tian{27.9\%}, 31.8\%, and 31.0\%, respectively.
These results demonstrate that the skills produced by \tool{} help the agent reveal substantially more real bugs while achieving higher precision.

Human-written skills provide limited benefits and do not consistently improve different base agents.
On OpenHands, human skills improve F1 from 0.08 to 0.16.
In contrast, on mini-SWE-agent, the same skills slightly reduce F1 from 0.28 to 0.26.
This is because human-written skills mainly provide generic testing heuristics, such as exploring boundary conditions and mocking dependencies.
Such broad testing heuristics can help compensate for missing testing guidance in weaker agents, but are less effective when the base agent already has stronger testing capability.
Furthermore, existing evolution methods also provide limited gains.
On OpenHands, Live-SWE-agent and Trace2Skill improve F1 from 0.08 to 0.22 and 0.19, respectively.
On mini-SWE-agent, they improve F1 only slightly from 0.28 to 0.29.
This is mainly because both of them perform evolution through local updates, making it difficult to coordinate them with existing skills or verify whether they remain useful beyond the cases that produced them.
In contrast, \tool{} co-evolves inter-skill relationships via the SRG and generalizes local updates into reusable capabilities, enabling robust bug-triggering and test-construction knowledge while avoiding overfitting and regressions.

\begin{figure}[t]
  \centering
  \includegraphics[width=0.95\linewidth]{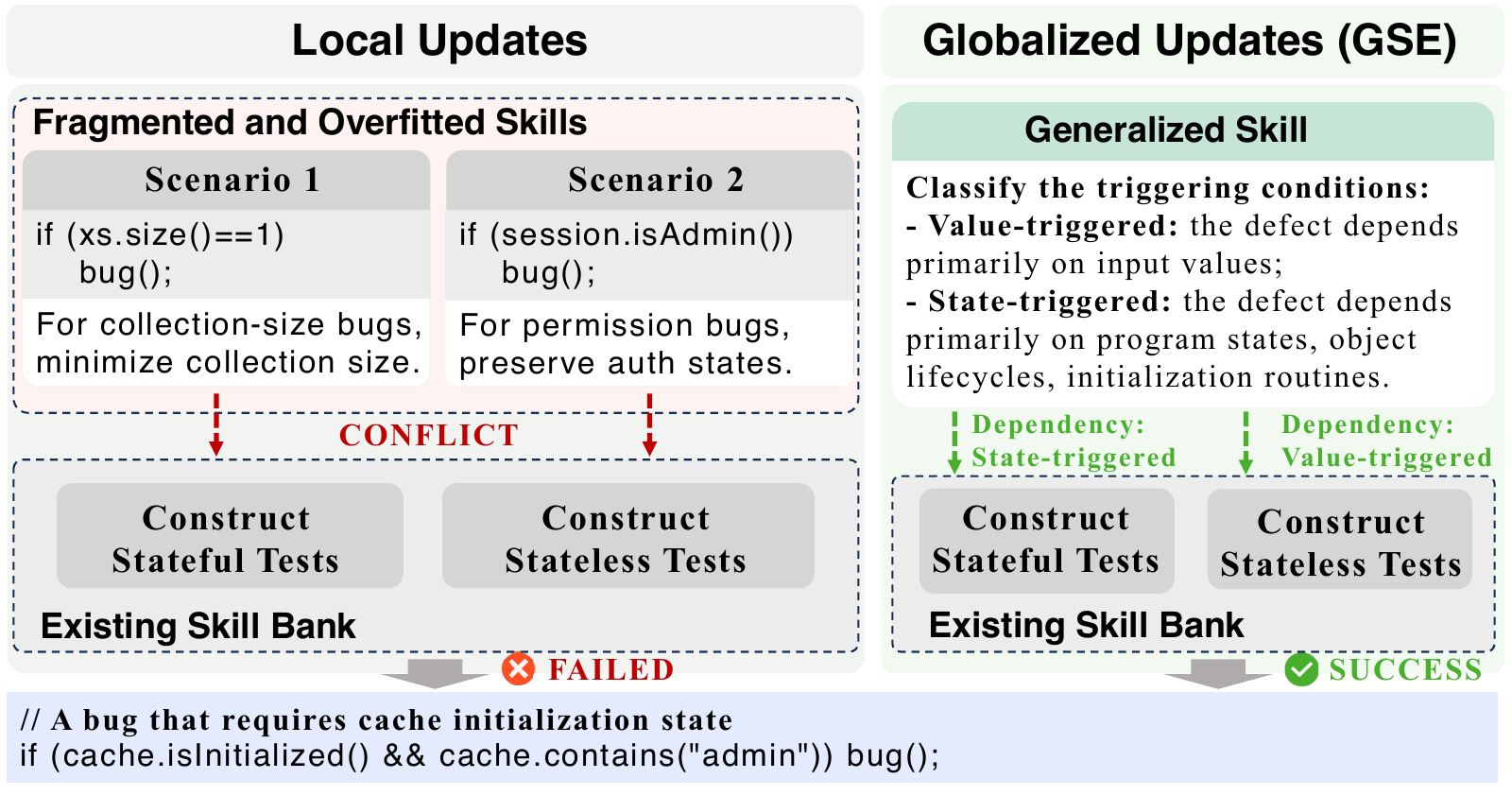}
  \caption{A Case Study for Bug-Triggering Test Generation}
  \label{fig:case1}
\end{figure}

\smallskip
\noindent
\textbf{Case Study.}
Figure~\ref{fig:case1} illustrates a simplified example comparing local updates with our globalized approach. 
Under a local-update baseline, the agent learns isolated rules for specific scenarios, such as minimizing collections for \texttt{collection-size} bugs, or preserving states for \texttt{permission} bugs. 
As depicted on the left, these specific updates not only overfit to their local execution contexts but also introduce \textit{conflicts} within the skill bank by redundantly embedding custom test-construction logic that clashes with existing skills. 
Consequently, when encountering a new defect requiring a cache initialization state (\texttt{cache.isInitialized()}), the agent fails to retrieve any relevant skill due to these overly rigid applicability conditions.
\yc{
In contrast, as shown on the right, \tool{} addresses these issues. Through cluster-based consolidation, \tool{} abstracts the fragmented local updates into a unified, higher-level skill: \textit{Determine Bug Triggering Mechanisms}, which classifies defects as either \textit{value-triggered} or \textit{state-triggered}. 
More importantly, guided by the Skill Relation Graph (SRG), \tool{} resolves the previous structural conflicts by establishing a formal \textit{dependency} relationship between this newly generalized diagnostic skill and the existing \textit{Construct Stateful/Stateless Tests} skills. 
As a result, when applied to the unseen cache-state bug, the agent successfully leverages the generalized skill to classify the defect as state-triggered, and subsequently follows the SRG dependency to invoke the appropriate stateful test-construction skill, effectively exposing the bug.
}

\begin{tcolorbox}[colback=gray!5]
\textbf{RQ1 Summary}: 
\yc{
\tool{} achieves the best performance on bug-revealing test generation, improving F1-score by 31.0\%$\sim$93.8\%.
}
This demonstrates that globalized skill evolution can effectively improve the agent's ability to generate bug-revealing tests
\end{tcolorbox}

\subsection{RQ2: Effectiveness on False-Positive Bug Report Filtering}
\subsubsection{Process}
For each report, the agent determines whether it corresponds to a bug or a false alarm.
We compare the prediction against the developer-confirmed label and report precision, recall, and F1-score.

\subsubsection{Results}
\begin{table}[t]
\centering
\caption{Effectiveness Comparison on False-Positive Filtering}
\label{tab:rq2}
\resizebox{0.75\columnwidth}{!}{
\begin{tabular}{l |ccc| ccc}
\toprule
\multirow{2}{*}{\textbf{Method}} & \multicolumn{3}{c|}{\textbf{OpenHands}} & \multicolumn{3}{c}{\textbf{mini-SWE-agent}} \\
& \textbf{Precision} & \textbf{Recall} & \textbf{F1} & \textbf{Precision} & \textbf{Recall} & \textbf{F1} \\
\midrule
Base Agent & 0.25 & 0.73 & 0.37 & 0.23 & 0.78 & 0.36 \\
+ Human Skill & 0.29 & 0.80 & 0.42 & 0.25 & 0.81 & 0.38 \\
+ Live-SWE-agent & 0.28 & 0.82 & 0.42 & 0.26 & 0.83 & 0.39 \\
+ Trace2Skill & 0.29 & 0.84 & 0.43 & 0.25 & 0.84 & 0.39 \\
\midrule
+ \tool{} & \textbf{0.55} & \textbf{0.95} & \textbf{0.70} & \textbf{0.30} & \textbf{0.97} & \textbf{0.46} \\
\bottomrule
\end{tabular}
}
\end{table}

Table~\ref{tab:rq2} presents the results for false-positive filtering.
On OpenHands, \tool{} achieves a precision of 0.55, recall of 0.95, and F1-score of 0.70, substantially outperforming the strongest baselines, which achieve only 0.29 precision, \tian{0.84} recall, and 0.43 F1-score.
On mini-SWE-agent, \tool{} improves precision from 0.26 to 0.30, recall from 0.84 to 0.97, and F1-score from 0.39 to 0.46 over the strongest baselines.
Across both agents, \tool{} improves precision by \tian{15.4\%$\sim$96.4\%}, recall by \tian{13.1\%$\sim$19.8\%}, and F1 by \tian{17.9\%$\sim$66.7\%}.
These results demonstrate that globalized skill evolution not only helps the agent identify more real bugs, but also substantially reduces false alarms.

Human-written skills provide only modest improvements, increasing precision from 0.25 to 0.29 and F1 from 0.37 to 0.42 on OpenHands, and precision from 0.23 to 0.25 and F1 from 0.36 to 0.38 on mini-SWE-agent.
This is because false-positive filtering in enterprise environments requires reasoning about complex business logic, API contracts, and project-specific engineering conventions.
Manually crafted heuristics are difficult to adapt to the diverse business contexts and industrial repositories.
Without such domain knowledge, the agents are easily misled by superficial code patterns.
\yc{
Existing evolution methods improve recall but remain limited in precision.
On OpenHands, Live-SWE-agent and Trace2Skill increase recall from 0.73 to 0.82 and \tian{0.84}, respectively, while on mini-SWE-agent they achieve recalls of 0.83 and 0.84.
However, their precision ranges from only 0.25 to 0.29.
This is because both methods rely on local updates and do not explicitly verify whether the learned behaviors generalize across cases.
In contrast, \tool{} achieves much higher precision while preserving high recall, demonstrating its effectiveness.
}

\smallskip
\noindent
\textbf{Case Study.}
We briefly summarize a representative case here.
Both Live-SWE-agent and Trace2Skill learned multiple project-specific nullability skills, some treating unchecked dereferences as real bugs and others treating them as benign due to framework contracts or upstream validation. These overfitted and conflicting skills made downstream bug validation unreliable. In contrast, \tool{} generalized them into a higher-level capability for reasoning about nullability guarantees and propagated the resulting abstraction to related bug-validation skills through the SRG.

\begin{tcolorbox}[colback=gray!5]
\textbf{RQ2 Summary}: \yc{\tool{} achieves the best performance on false-positive filtering, improving precision by \tian{15.4\%$\sim$96.4\%} and recall by \tian{13.1\%$\sim$19.8\%.}}
This demonstrates that globalized skill evolution can improve the agent's ability to identify false alarms.

\end{tcolorbox}

\subsection{RQ3: Ablation Study}
\subsubsection{Process}
In this RQ, we investigate the contributions of the two key components of \tool{}, namely the Skill Relation Graph (SRG) and the skill generalization mechanism. To isolate their effects, we construct two variants:
\begin{itemize}
\item \textbf{\nograph{}} removes the SRG and replaces relationship-aware skill evolution with independent skill updates, disabling both inter-skill relationship evolution and propagation.

\item \textbf{\nogen{}} removes the global skill generalization mechanism. Evolution proposals are directly merged into the skill bank after local validation, without cluster-based consolidation or replay-driven validation.

\end{itemize}
To balance evaluation cost and generalizability, we use OpenHands as the representative base agent for both tasks.

\subsubsection{Results}

Table~\ref{tab:rq3} shows that removing either component substantially degrades performance, demonstrating that both relationship-aware evolution and skill generalization are essential for maintaining an effective skill bank.
Removing the Skill Relation Graph (\textbf{\nograph{}}) consistently reduces performance, with F1 dropping from 0.31 to 0.27 on bug-revealing test generation and from 0.70 to 0.59 on false-positive filtering. 
This result indicates that explicitly modeling inter-skill relationships is critical for preserving the global consistency of the skill bank. 
Without the SRG, evolution proposals are generated independently and may invalidate assumptions relied upon by related skills. 
For example, updating a repository exploration skill to aggressively truncate file outputs may simultaneously break downstream analysis skills that depend on complete file contents. 

Removing the skill generalization mechanism (\textbf{\nogen{}}) also degrades performance, with F1 dropping from 0.31 to 0.27 on test generation and from 0.70 to 0.54 on false-positive filtering. 
This finding suggests that local effectiveness alone is insufficient for long-term skill evolution. 
Without clustering, capability abstraction, and replay-driven validation, the system directly incorporates case-specific evolution proposals into the skill bank, causing it to accumulate fragmented and overfitted knowledge. 
In practice, such updates often solve the originating failure but introduce regressions on related tasks. 
Overall, the full \tool{} achieves the best performance, confirming that effective skill evolution requires both preserving global consistency and learning reusable capabilities.

\begin{table}[t]
\centering
\caption{Ablation Study}
\label{tab:rq3}
\resizebox{0.7\columnwidth}{!}{
\begin{tabular}{l|ccc|ccc}
\toprule
\multirow{2}{*}{\textbf{Method}} & \multicolumn{3}{c|}{\textbf{Test Generation}} & \multicolumn{3}{c}{\textbf{False-Positive Filtering}} \\
& \textbf{Precision} & \textbf{Recall} & \textbf{F1} & \textbf{Precision} & \textbf{Recall} & \textbf{F1} \\
\midrule
\nograph{} & 0.31 & 0.24 & 0.27 & 0.44 & 0.89 & 0.59 \\
\nogen{} & 0.32 & 0.23 & 0.27 & 0.39 & 0.91 & 0.54 \\
\midrule
\tool{} & \textbf{0.35} & \textbf{0.28} & \textbf{0.31} & \textbf{0.55} & \textbf{0.95} & \textbf{0.70} \\
\bottomrule
\end{tabular}
}
\end{table}

\begin{tcolorbox}[colback=gray!5]
\textbf{RQ3 Summary}: Both the skill relation graph and the skill generalization are crucial to \tool{}'s effectiveness.
\end{tcolorbox}

\subsection{RQ4: Cost Analysis}
\subsubsection{Process}
We further measure the token cost during both skill evolution and downstream task execution. 
We compute the mean token consumption per case and macro-average the results across the two tasks and two agents. 

\begin{table}[t]
\caption{Token consumption macro-averaged across tasks and agents.}
\label{tab:token_cost}
\centering
\resizebox{0.7\linewidth}{!}{
\begin{tabular}{lrrrc}
\toprule
\textbf{Method} & \textbf{Input} & \textbf{Output} & \textbf{Total} & \textbf{$\Delta$ Total} \\
\midrule
\multicolumn{5}{l}{\textit{(a) Evolution process per case}} \\
Trace2Skill & 335.52K & 22.12K & 357.64K & -- \\
\tool{}     & 388.79K & 12.76K & 401.55K & \textcolor{red}{$+12.28\%$} \\
\midrule
\multicolumn{5}{l}{\textit{(b) Task execution per case}} \\
Base Agent       & 483.94K & 5.36K & 489.31K & -- \\
+ Human Skills   & 614.74K & 5.84K & 620.58K & \textcolor{red}{$+26.8\%$} \\
+ Live-SWE-agent & 608.07K & 5.54K & 613.61K & \textcolor{red}{$+25.4\%$} \\
+ Trace2Skill    & 613.95K & 5.70K & 619.65K & \textcolor{red}{$+26.6\%$} \\
+ \tool{}        & 588.29K & 4.92K & 593.21K & \textcolor{green}{$+21.2\%$} \\
\bottomrule
\end{tabular}
}
\end{table}

\subsubsection{Results}
Table~\ref{tab:token_cost} presents the token cost of skill evolution and downstream task execution.
The results show that \tool{} achieves substantially better effectiveness with modest additional cost.
For skill evolution, \tool{} consumes an average of \textit{401.55K} tokens per case, compared with \textit{357.64K} for Trace2Skill, an increase of \textit{12.28\%}. 
We exclude Live-SWE-agent because its online evolution process is intertwined with task execution, making the evolution cost inseparable from task-solving cost.
For downstream execution, agents enhanced with \tool{} consume \textit{593.21K} tokens per case, compared with \textit{489.31K}, \textit{620.58K}, \textit{613.61K}, and \textit{619.65K} for the base agent, human skills, Live-SWE-agent, and Trace2Skill, respectively. Although \tool{} introduces a moderate overhead relative to the base agent, it requires fewer tokens than all other methods, reducing token consumption by \textit{4.4\%}, \textit{3.3\%}, and \textit{4.3\%} compared with human skills, Live-SWE-agent, and Trace2Skill, respectively. This finding suggests that the additional prompt budget introduced by evolved skills can be offset by more effective reasoning and exploration strategies.

\begin{tcolorbox}[colback=gray!5]
\textbf{RQ4 Summary}: \tool{} incurs a modest additional evolution cost while requiring the fewest execution-time tokens among all methods, showing that globalized skill evolution can substantially improve effectiveness without prohibitive computational overhead.
\end{tcolorbox}

\section{Industrial Deployment}
\yc{
A potential concern is whether globalized skill evolution remains beneficial when applied to stronger coding agents that already incorporate domain-specific knowledge and optimized reasoning workflows.
To investigate this question, we instantiate \tool{} on our industrial partner's proprietary bug-triage agent and evaluate it on the false-positive filtering task.
Unlike OpenHands and mini-SWE-agent, this internal agent is specifically designed for enterprise bug analysis and incorporates customized tools, proprietary APIs, and domain-specific prompting strategies.
Its base performance already exceeds that of both open-source agents (F1-score: 0.43 vs. 0.37/0.36), making it a representative strong industrial agent.
}

\yc{
Table~\ref{tab:internal_agent} presents the results.
The base internal agent achieves a precision of 0.29, recall of 0.83, and F1-score of 0.43.
Human-written skills slightly improves performance to 0.30 precision, 0.83 recall, and 0.44 F1-score.
In contrast, \tool{} achieves a precision of 0.56, recall of 0.97, and F1-score of 0.71.
Compared with the base internal agent, \tool{} improves precision by 93.1\%, recall by 16.9\%, and F1-score by 65.1\%.
Compared with human-written skills, \tool{} improves precision by 86.7\%, recall by 16.9\%, and F1-score by 61.4\%.
These results suggest that the benefits of globalized skill evolution are not limited to relatively lightweight open-source agents.
Even when applied to a stronger industrial agent equipped with domain-specific knowledge and optimized reasoning capabilities, \tool{} continues to produce substantial improvements.
}

\begin{table}[t]
\caption{Generalizability to the industrial internal agent.}
\label{tab:internal_agent}
\centering
\begin{tabular}{l|ccc}
\toprule
\textbf{Method} & \textbf{Precision} & \textbf{Recall} & \textbf{F1} \\
\midrule
Internal Agent                & 0.29 & 0.83 & 0.43 \\
+ Human Skills & 0.30 & 0.83 & 0.44 \\
\midrule
+ \tool{}                   & \textbf{0.56} & \textbf{0.97} & \textbf{0.71} \\
\bottomrule
\end{tabular}
\end{table}

\section{Threats to Validity}
\label{sec:threats}

The \textbf{external threats} primarily stem from the generalizability of \tool{} across coding agents, programming languages, and software engineering tasks.
\yc{
To mitigate this threat, we evaluate \tool{} on two state-of-the-art open-source coding agents (i.e., OpenHands and mini-SWE-agent) and further deploy it on our industrial partner's proprietary agent.
Regarding tasks, we leverage two representative software engineering tasks, bug-revealing test generation and false-positive bug report filtering, which jointly cover several core capabilities required by software engineering agents.
}
Our evaluation also spans both Java and Go across diverse domains and scales.
The consistent improvements confirm that \tool{} generalizes beyond a particular benchmark, partially mitigating this threat.

The \textbf{internal threats} primarily stem from potential implementation errors in \tool{} or the compared techniques.
To mitigate this risk, we used the officially maintained OpenHands~\cite{openhands} and mini-SWE-agent~\cite{swe_agent} as base agents.
For the compared techniques (i.e., Live-SWE-agent and Trace2Skill), we used their released artifacts and strictly followed the procedures described in their papers.
The implementation of \tool{} was carefully reviewed and tested by the authors.

The \textbf{construct threats} include LLM randomness and data leakage.
To reduce randomness and ensure deterministic evaluation, we set the LLM temperature to zero for all techniques, following common practice in prior LLM-based studies~\cite{rted,sega, fuzz4all}.
To mitigate data leakage, we adopt a one-project-held-out evaluation protocol for both tasks. This ensures that no task from the held-out project is used to evolve the skill bank, and tests whether the evolved skills generalize across projects. For the bug-revealing test generation task, we also follow existing work~\cite{auger, rted} and check whether any bug-triggering tests generated by \tool{} match the benchmark reference tests. None of the generated tests align with the reference tests. For false-positive filtering, IndustrialBugs consists entirely of proprietary codebases, and our industrial partner confirmed that these codebases are excluded from the training corpora of public LLMs. Moreover, all baselines use the same underlying LLM and agent harnesses, so performance improvements are not due to the data leakage of LLMs.

\section{Related Work}
\label{sec:related_work}

\smallskip
\noindent
\textbf{Skills for LLM Agents.}
LLM agents increasingly rely on externalized knowledge to improve behavior without updating model parameters.
Early work shows that agents can benefit from tool-use learning, verbal reflection, and automatically generated workflows~\cite{toolformer,reflexion,voyager,aflow}.
Recent efforts further formalize this paradigm into comprehensive skill systems, where skills are treated as explicit artifacts that encapsulate reusable reasoning strategies, decision procedures, and task-specific knowledge for on-demand use~\cite{xu2026agent}.
Large-scale evaluations show both the promise and fragility of this paradigm.
SkillsBench~\cite{skillsbench} finds that well-designed skills can improve agents across diverse professional tasks, whereas SWE-Skills-Bench~\cite{sweskills} reports that software engineering skills provide uneven gains and may even degrade performance when their domain fit or contextual compatibility are mismatched with the target task.
To better support large-scale skill banks, recent work has explored skill retrieval and routing techniques. 
For example, SkillRet~\cite{skillret} studies retrieval over large-scale skill banks, while SkillRouter~\cite{skillrouter} investigates efficient skill selection.
\textit{In contrast to prior work on static skills, \tool{} continuously evolves an interconnected skill bank through relation-aware evolution and skill generalization.}

\smallskip
\noindent
\textbf{Agent Evolution.}
Agent evolution aims to improve an agent based on its execution.
Existing approaches can be categorized into two lines.
The \textit{first line} improves the agent itself through parameter updates.
Recent agentic reinforcement learning methods optimize agent policies for long-horizon reasoning and interaction tasks, such as ReTool~\cite{retool}, WebRL~\cite{webrl}, and OpenClaw-RL~\cite{openclawrl}, which train agents to better utilize tools and interact with environments through reinforcement learning.
While these methods can substantially improve agent capabilities, they require expensive model training.
The \textit{second line} keeps the underlying model fixed and externalizes experience into reusable artifacts.
One direction evolves the agent tools or execution workflow, such as AFlow~\cite{aflow}, Meta-Harness~\cite{metaharness}, and Live-SWE-agent~\cite{liveswe}.
Another direction maintains reusable memories or experiences, including reflection-based methods such as Reflexion and ExpeL~\cite{reflexion,expel}, and persistent memory systems such as ReasoningBank and Memento~\cite{reasoningbank,memento}.
Recently, increasing attention has shifted to skill evolution due to its lightweight, interpretable, and highly reusable nature.
Accordingly, a third direction packages reusable experience into explicit skills, with recent work studying how to extract, revise, and organize skills from execution trajectories and feedback~\cite{trace2skill,codeskill,skillrl,skill1}.
However, existing methods typically perform skill evolution through isolated local updates and do not consider their global impact.
\textit{In contrast, \tool{} performs globalized skill evolution by jointly evolving inter-skill relationships and generalizing locally effective updates into reusable skills.}

\section{Conclusion}

In this paper, we propose \tool{}, a globalized skill evolution framework that formulates skill evolution as a global optimization problem over an interconnected and reusable skill bank. 
By co-evolving inter-skill relationships through a Skill Relation Graph (SRG) and generalizing locally effective evolution proposals into reusable skills through cluster-based consolidation and replay-driven validation, \tool{} preserves global consistency while preventing overfitting and behavioral regressions. 
Our extensive evaluation demonstrates that \tool{} consistently outperforms existing techniques across diverse scenarios. In particular, it improves precision by \tian{6.1\%$\sim$34.1\%} and recall by \tian{31.8\%$\sim$180.0\%} for bug-revealing test generation. For false-positive bug report filtering, \tool{} achieves improvements of \tian{15.4\%$\sim$96.4\%} in precision and \tian{13.1\%$\sim$19.8\%} in recall, confirming the effectiveness of globalized skill evolution.

\bibliographystyle{plain}
\bibliography{references}

\end{document}